\documentclass{ws-procs9x6}

\begin{document}

\title{Exclusively produced $\rho^0$ asymmetries on the deuteron and future GPD
measurements at COMPASS}

\author{C. Schill on behalf of the COMPASS collaboration}

\address{Physikalisches Institut der Albert-Ludwigs-Universit\"at Freiburg, \\
Hermann-Herder Str.3, D-79104 Freiburg, Germany.\\
E-mail: Christian.Schill@cern.ch}

\begin{abstract} Generalized parton distributions (GPDs) provide a new and
powerful framework for a complete description of  the nucleon structure. They
can provide a three-dimensional picture of how the quarks and gluons form a
nucleon. GPDs can be probed experimentally in hard exclusive meson production 
or deeply virtual Compton scattering (DVCS).  The COMPASS experiment at CERN
is a unique place to study these reactions. At COMPASS, a high energy polarized
positive or negative muon beam is scattered off a polarized or unpolarized
fixed target. First results for exclusive $\rho^0$ meson
production are shown. The transverse target spin asymmetry for exclusively
produced $\rho^0$ on a transversely polarized deuteron target has been
measured. Prospects for future measurements of DVCS and exclusive meson
production at COMPASS will be shown. The experiment will use the existing
COMPASS spectrometer with a new target, a new recoil detector and extended
calorimetry. Simulations for different models and a test of the recoil detector
have been performed.

\end{abstract}


\bodymatter

\section{Introduction}

Generalized parton distributions (GPDs) have been introduced about 10 years ago
\cite{Mueller,Radyushkin,Ji} and provide a comprehensive description of the nucleon. They can
describe elastic form factors and polarized and unpolarized parton
distributions at the same time. In addition, they may provide a handle on the
orbital angular momentum of the quarks in the nucleon via the Ji sum rule \cite{Ji}.
GPDs may be accessed experimentally by measuring cross-sections and
asymmetries in deeply virtual Compton scattering and exclusive meson production
processes at COMPASS. 

GPDs describe the quantum-mechanical amplitude for "taking out" a parton (quark
or gluon) from a fast moving nucleon and "putting it back" with a different
momentum, leading to a small momentum transfer to the nucleon. The GPDs are
functions of three parameters, the longitudinal quark momentum fraction $x$, the
longitudinal momentum transfer to the nucleon $\xi=x_{bj}/(2-x_{bj})$ and the
squared momentum transfer to the target nucleon $t$. The latter is the Fourier
conjugate of the transverse impact parameter $r$ of the scattering. 

At leading twist four GPDs are necessary to describe the quark structure of the
proton. The GPDs $H$ and $\tilde{H}$ conserve the helicity of the proton and
are generalizations of the ordinary parton distributions measured in DIS. In
the forward limit $\xi=0$ and $t=0$ the unpolarized GPD $H$ reproduces the
unpolarized quark distribution $q(x)$. The polarized GPD $\tilde{H}$ gives the polarized
quark distribution $\Delta q(x)$.

The GPDs $E$ and $\tilde{E}$ allow a proton helicity flip and imply a transfer
of orbital angular momentum to the nucleon. This is only possible for a non-zero
transverse momentum transfer $t$, which is new with respect to ordinary parton
distribution functions.

\section{The COMPASS experiment}

The COMPASS experiment is a fixed target experiment at the SPS accelerator at
CERN. It scatters a longitudinally polarized muon beam on a polarized solid
state target. From 2002 to 2004, COMPASS took data with a muon beam of
$160$~GeV and a longitudinally or transversely polarized $^6LiD$ target. For
about $20$\% of the beam time, the target nucleons have been polarized 
transversely with respect to the beam polarization. The data presented here for 
the exclusive $\rho^0$ production have been taken with a transverse
polarization of the target of about $50$\%. 

The scattered muon and produced hadrons are detected in a two-stage 
spectrometer \cite{NIM}, which provides high momentum resolution and excellent
particle
identification.

\section{Exclusive $\rho^0$ production}

One observable sensitive to the GPDs $E$ and $H$ is the transverse target spin
asymmetry in the exclusive $\rho^0$ vector meson production. The reaction
studied is

\begin{equation*}
\mu N \rightarrow \mu´ \rho^0 N,
\end{equation*}

where $N$ is a quasi-free nucleon in the polarized deuteron target. The
reaction can be described as virtual photoproduction of a $\rho^0$:
$\gamma^*N\rightarrow \rho^0N$. To interpret the results in terms of GPDs, the
following conditions have to be fulfilled: The virtual photon has to be
longitudinally polarized, since only for longitudinal photons factorization
between the hard reaction and the soft GPD part has been shown. The negative
four-momentum transfer squared of the virtual photon $Q^2$ has to be large
\cite{rho-paper} and the momentum transfer squared $t$ between initial and
final nucleon has to be small.

At COMPASS, the outgoing $\rho^0$ and the scattered muon is detected, but the
recoil proton is not. The $\rho^0$ is reconstructed from the invariant mass of
the two pions by applying an invariant mass cut of
$|M_{\pi^+\pi^-}-M_\rho|<0.3$~GeV. The exclusivity is insured by a cut on the
missing energy of $|E_{miss}|<2.5$~GeV. To further minimize
the non-exclusive background, a cut on the transverse momentum of the $\rho^0$
of $p_T^2<0.5$~(GeV/c)$^2$ has been applied. For a reliable definition of the
azimuthal angle of the produced $\rho^0$ with respect to the lepton scattering
plane, a lower limit has been imposed on the transverse momentum
$p_T^2>0.01$~(GeV/c)$^2$.

Deep-inelastic scattering events were selected by a cut on the negative
four-momentum transfer $Q^2>1$~(GeV/c)$^2$ and on the invariant mass of the
hadronic system $W>5.0$~GeV/c. The energy fraction of the virtual photon with
respect to the beam momentum is limited in the range $0.1 < y < 0.9$. After all
cuts $270k$ exclusively produced $\rho^0$ mesons have been obtained in total. 

\begin{figure}
\centerline{\hspace*{2cm}
\includegraphics[clip, bb=100 280 642 585,width=0.65\textwidth]{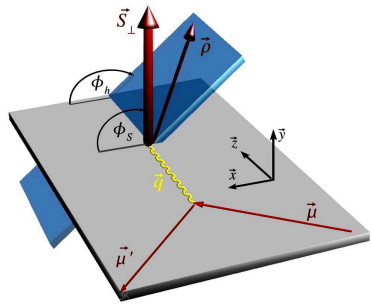}}
\caption{Definition of $\phi_h$ and $\phi_S$ for exclusive $\rho^0$ production.}
\label{fig1}
\end{figure}

The azimuthal target spin asymmetry has been evaluated with respect to the
angle ($\phi_h-\phi_S$) (Fig. \ref{fig1}), where $\phi_h$ is the azimuthal angle
of the produced $\rho^0$ meson with respect to the lepton scattering plane and
$\phi_S$ the azimuthal angle of the target spin vector.

The transversely polarized target consists of  two target cells with opposite
polarization direction. Their polarization is reversed about once per week. To
minimize systematic effects, the asymmetry is extracted from the counting rates
in both cells and both target configurations with the double ratio method
described in \cite{Compass}. The extracted raw asymmetry $\epsilon$ is
normalized to the target polarization $\left<P\right>$ and the dilution factor $f=0.38$ of
the target:
\begin{equation}
A_{UT}=\frac{\epsilon}{f\cdot \left<P\right>}
\end{equation}

\section{Results for exclusive $\rho^0$ production}

\begin{figure}\hspace*{-0.7cm}
\includegraphics[width=0.55\textwidth]{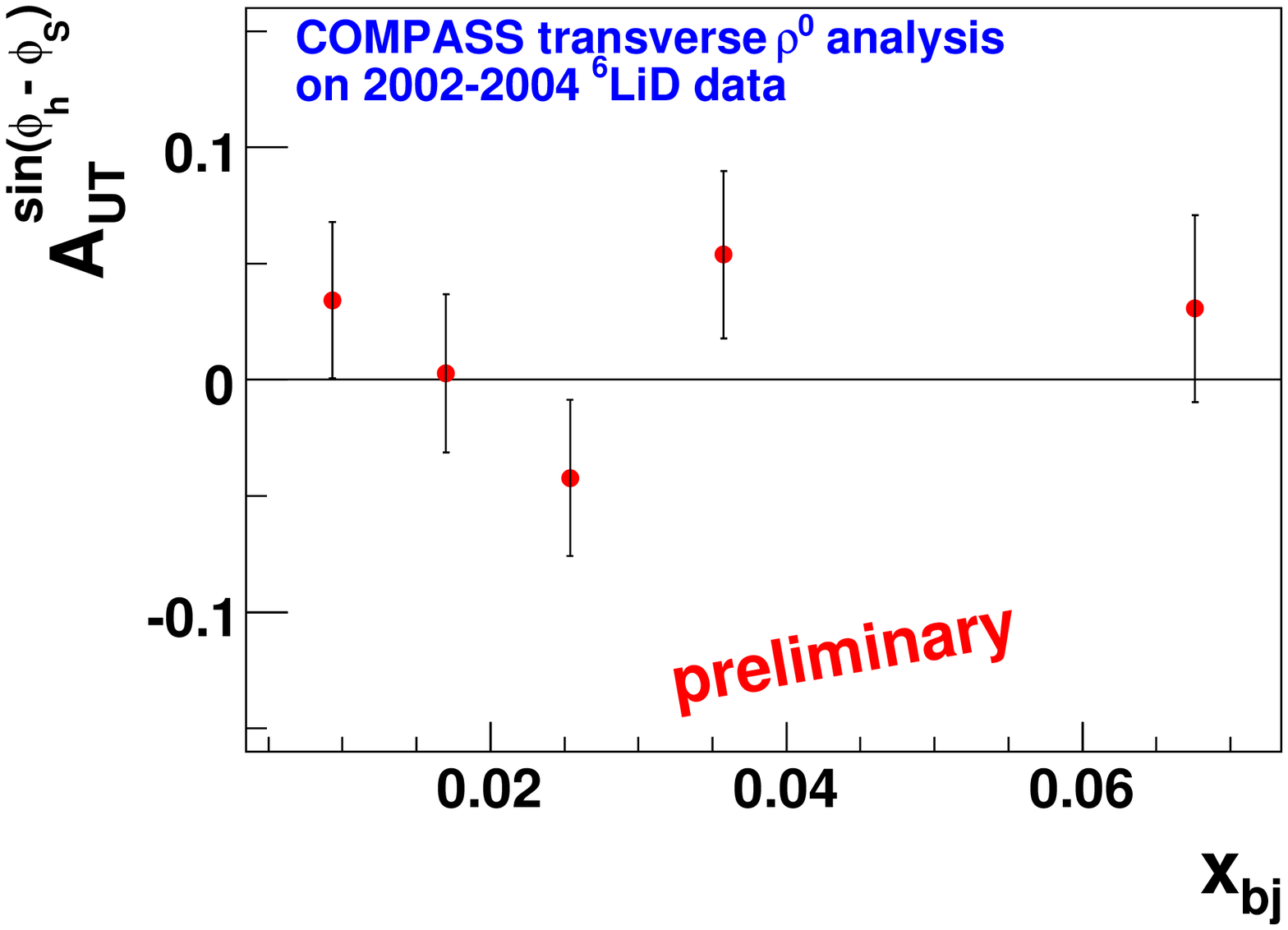}\hspace*{-0.2cm}
\includegraphics[width=0.55\textwidth]{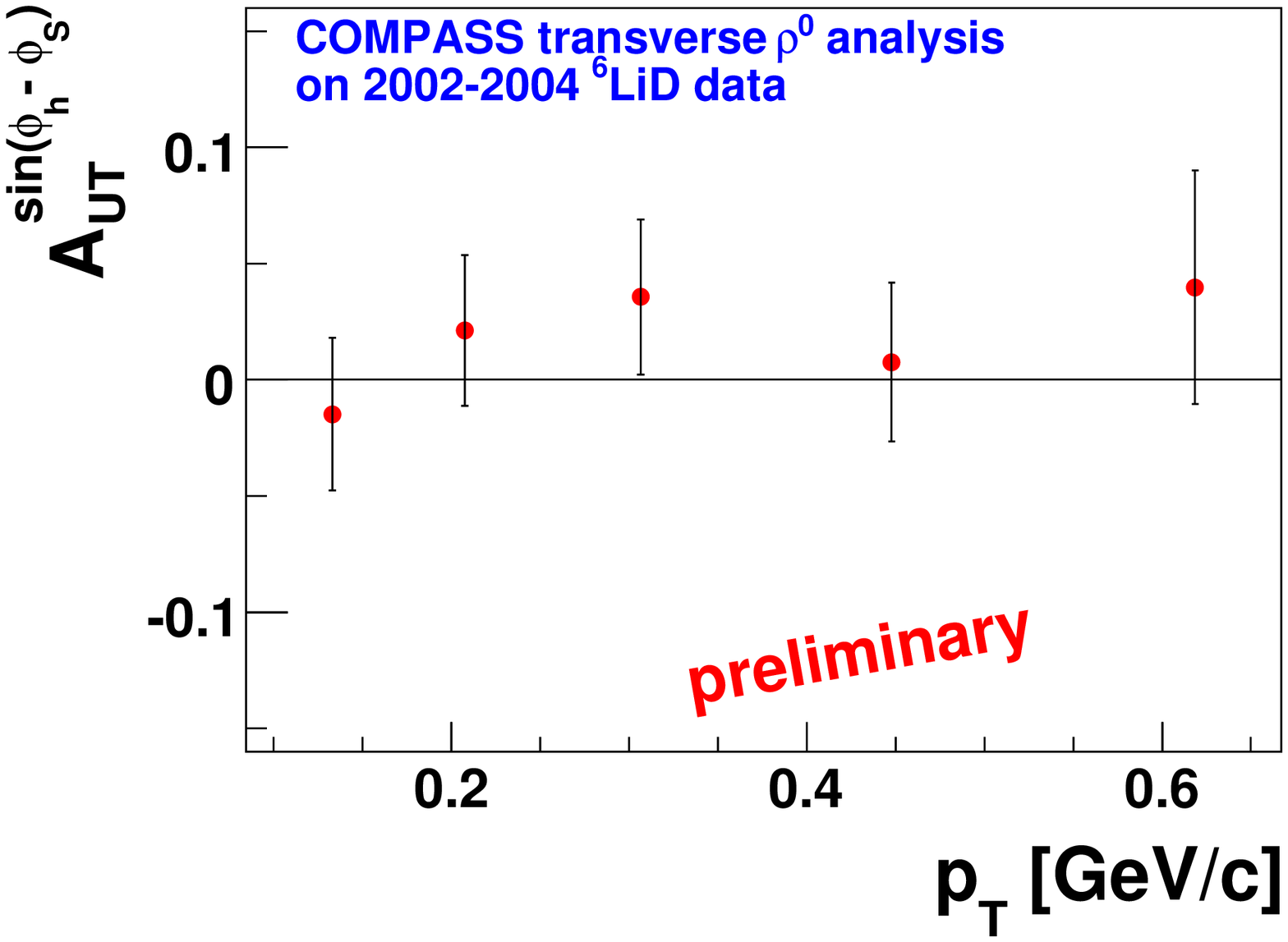}\hspace*{-0.5cm}
\caption{Transverse $\rho^0$ target spin asymmetries as a function of 
$x_{bj}$ (left) and $p_T$ (right).}
\label{Results}
\end{figure}

The transverse target spin asymmetry $A_{UT}$ in exclusive $\rho^0$ production
is plotted in Fig.~\ref{Results} as a function of $x_{bj}$ and $p_T$. 
The measured asymmetries are small and within their statistical precision
compatible with zero. Further analysis will allow to disentangle the
contribution of the incoherent production on a quasi-free nucleon from the
coherent production on the deuteron. Both mechanisms can be distinguished by 
the produced transverse momentum, since coherent production happens preferable
at lower transverse momenta.

For a comparison to GPD model calculations, the asymmetry for  longitudinal
photons has to be extracted \cite{rho-paper}. A method to extract the asymmetry
for longitudinal photons was proposed in \cite{Diehl}: the contributions of
longitudinal and transverse $\rho^0$ can be estimated from the angular
distribution of the  decay products. By assuming s-channel helicity
conservation, we can determine the contribution from longitudinal photons.

In addition, in 2007 we took data with a transversely polarized proton
target. Repeating the analysis on our proton data will provide new results on
$A_{UT}$ on the proton.

\section{Future GPD measurements at COMPASS}

\begin{figure}
\centerline{\includegraphics[width=\textwidth]{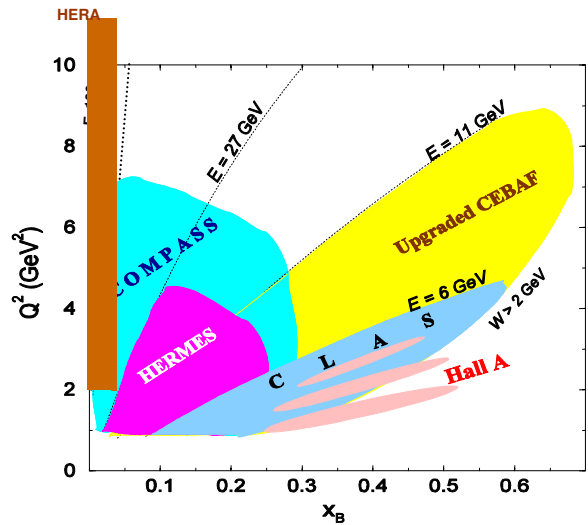}}
\caption{Kinematic coverage for various DVCS experiments.}
\label{Kinematics}
\end{figure}

The COMPASS experiment at CERN is a perfect place to study generalized parton 
distributions \cite{Proposal}. The GPDs will be accessed through deeply virtual
Compton scattering and hard exclusive meson production.  Due to the high energy
of the muon beam available at COMPASS  (between $100$ and $190$~GeV) the
kinematic range covered by the proposed GPD program  will be large enough to
provide a bridge between the HERA collider experiments \cite{H1, ZEUS} at very
small $x_{bj}$ and the JLAB \cite{CLAS,HALLA} and HERMES \cite{HERMES}
experiments at large $x_{bj}$ (Fig.~\ref{Kinematics}). 

The measurements of generalized parton distributions at COMPASS will be
performed with an unpolarized liquid hydrogen or deuterium target surrounded by
a recoil detector \cite{Proposal}. The CERN beam-line can provide both positive
and negative muons with opposite polarization of $80$\%. For beam charge
asymmetry measurements the beam charge can be changed once a day. The goal of
the experiment is to measure the cross section as well as spin and charge
asymmetries as a function of $x_{bj}$, $Q^2$ and $t$. At $190$~GeV, the DVCS
process is dominant over the competing Bethe-Heitler process and the cross
section can be measured. At $100$~GeV, the possibility to use muon beams of
opposite charge and polarization will allow to measure the charge and spin
asymmetries arising from the interference between DVCS and the Bethe-Heitler
process.  Measurements of the cross section difference of oppositely charged
and polarized muons give access to the real part of the DVCS amplitude, while
the cross section sum provides information about the imaginary part of the DVCS
amplitude \cite{DiehlMethod, Belitsky}. These observables have great
sensitivity to GPD models \cite{Guzey}. Hard exclusive meson production cross
sections will also be measured at the same time providing different constraints
on GPDs.

\section{Simulations} Simulations of the beam charge asymmetry for COMPASS have
been performed using three different model assumptions for the  DVCS cross section.
Two models are based on double distribution parameterizations of GPDs. Model~1
assumes $H(x, \xi,t)\propto q(x) F(t)$ \cite{Vanderhaegen}, while model~2
includes a correlation between $x$ and $t$: $H(x, \xi, t)\propto q(x)
e^{tb_\perp^2}$ with a parameter $b_\perp$ describing the transverse extension of the
partons. This ansatz reproduces the chiral quark soliton model \cite{Goeke}.
The model~3 uses a dual parametrization, where the $x,\xi$ dependence is
separated from the $\xi,t$ dependence. These calculations have been performed by
\cite{Guzey}.

\begin{figure} \centerline{\includegraphics[clip, bb= 100 240 495 647,
width=0.9\textwidth]{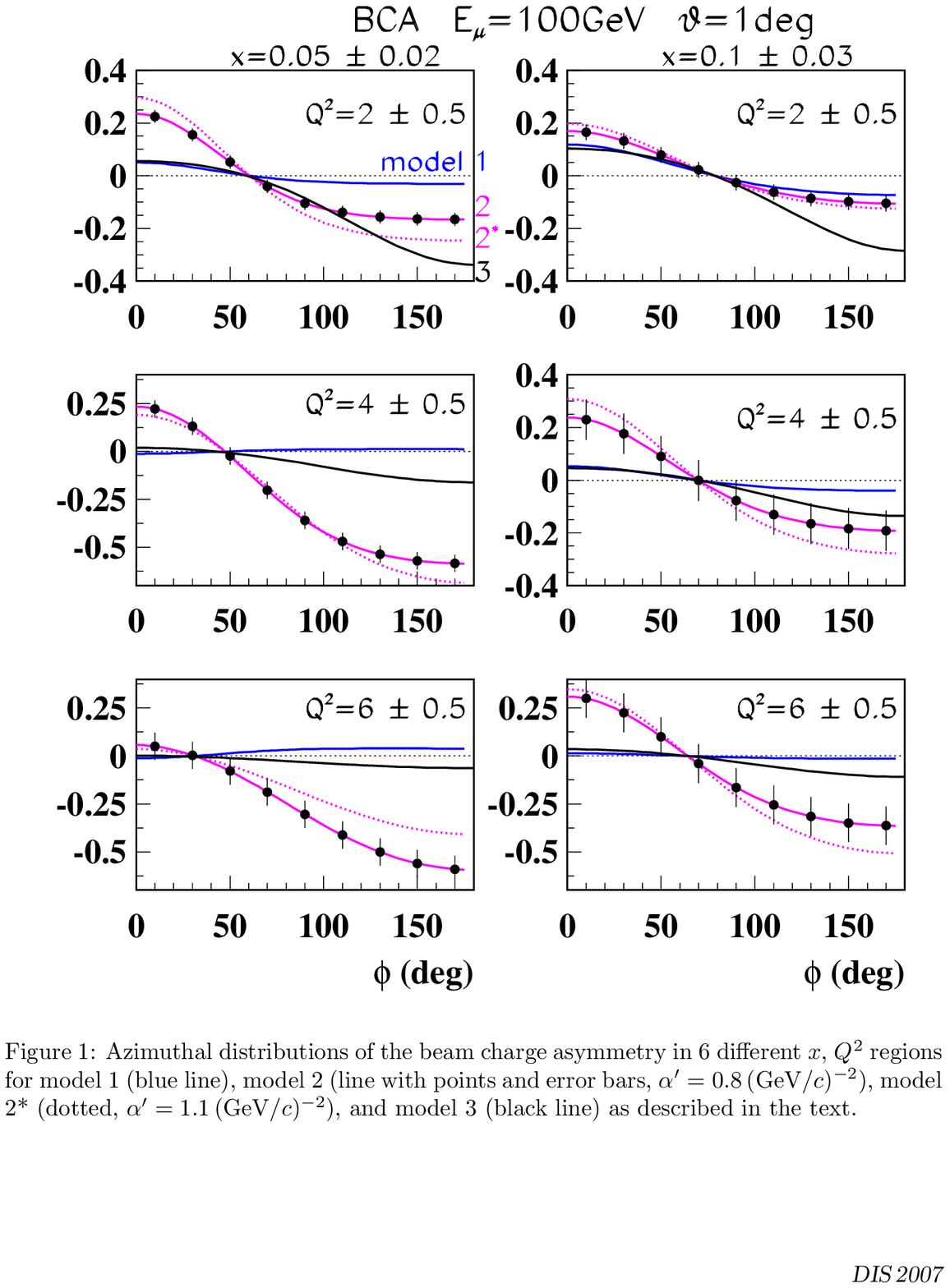}} \caption{Projected error bars for a
measurement of the beam charge DVCS asymmetry as an example for these
simulations in two domains of $x_{bj}$
($0.05$ on left and $0.10$ on right) and 3 bins of $Q^2$ (2, 4, 6 (GeV/c)$^2$)
from top to bottom. Projected error bars are compared to 3 different GPD models
described in the text: model 1 (blue line), model 2 (line with points and error
bars), model 3 (black line).} \label{Simulation} \end{figure}

With a luminosity of $1.3\cdot 10^{32}$~cm$^{-2}$s$^{-1}$ and an global efficiency of $25$\% and a
running of $150$ days the statistical precision shown in Fig.~\ref{Simulation} can be achieved. It is
possible to split the data into $6$ bins in $Q^2$ from $1.5$~(GeV/c)$^2$ to $7.5$~(GeV/c)$^2$ and in
$3$~bins in $x_{bj}$ from $0.03$ to $0.27$. A separation between the different models can be achieved
as shown in Fig.~\ref{Simulation}.

\section{Experimental realization}

The GPD experiment will use the existing COMPASS spectrometer with several new
detectors added. The interaction will take place in a newly designed $2.5$~m
long liquid hydrogen or deuterium target. It will be surrounded by a recoil
detector for the detection of the recoiling particles and to insure exclusivity
of the reaction. It will use the time-of-flight technique to determine the
recoiling proton momentum with a few percent precision. The recoil detector
will consist of two concentric barrels of scintillators read at both sides. To
allow for $10$~bins in $t$, a minimum time resolution of $300$~ps is needed. A
full size prototype has been tested in the COMPASS experimental area in fall
2006 and achieved a time resolution of $310$~ps.

In case of DVCS, the photon will be detected in the existing forward
calorimeters, an additional wide angle calorimeter covering lab angles up to
$20$ degrees will be added to improve the acceptance at high $x_{bj}$ and to
reject the $\pi^0$ background.

\section{Outlook}

A big
advantage of studying GPDs at the COMPASS experiment is  that its kinematic
domain covers both the valence and the sea quark region in terms of $x_{bj}.$ 
Currently a proposal for GPD measurements at COMPASS is being prepared. In
2008-2010 the recoil detector, the liquid hydrogen target and the large angle
calorimeter are being constructed. Data taking could start with the full set-up
as early as 2010. Therefore first results are expected well before possible future
projects at the JLAB $12$~GeV upgrade and the FAIR project at the GSI.

\end{document}